\documentclass[11pt]{article}
\usepackage{amsfonts}
\usepackage{amsmath, amssymb, amsthm}
\usepackage{fancyhdr}
\usepackage{graphicx}
\usepackage{dcolumn}
\usepackage{bm}

\begin{document}
{\setlength{\oddsidemargin}{1.4in}
\setlength{\evensidemargin}{1.4in} } \baselineskip 0.50cm
\begin{center}
{\LARGE {\bf Charged black holes in Vaidya backgrounds: Hawking's Radiation}}
\end{center}
\begin{center}Ng. Ibohal and L. Kapil \\
Department of Mathematics, Manipur University,\\
Imphal 795003, Manipur, INDIA.\\
E-mail: ngibohal@iucaa.ernet.in\\ laishramkapil@gmail.com
\end{center}

\begin{abstract}
In this paper we propose a class of embedded  solutions of Einstein's field equations describing  non-rotating Reissner-Nordstrom-Vaidya and rotating Kerr-Newman-Vaidya black holes. The Reissner-Nordstrom-Vaidya is obtained by embedding Reissner-Nordstrom solution into non-rotating Vaidya. Similarly, we also find the Kerr-Newman-Vaidya black hole, when Kerr-Newman embeds into the rotating Vaidya solution. The Reissner-Nordstrom-Vaidya solution is type D whereas Kerr-Newman-Vaidya metric is algebraically special type II of Petrov classification of space-time. These embedded solutions can be expressed in Kerr-Schild ansatze on different backgrounds. The energy momentum tensors for both non-rotating as well as rotating embedded solutions satisfy the energy conservation equations which show that they are solutions of Einstein's field equations. The surface gravity, area, temperature and  entropy are also presented for each embedded black hole. It is observed that the area of the embedded black holes is greater than the sum of the areas of the individual ones.
By considering the charge to be a function of radial coordinate it is shown that there is a change in the masses of the variable charged black holes. If such radiation continues, the mass of the black hole will evaporate completely thereby forming `instantaneous' charged black holes and creating embedded `negative mass naked singularities' describing the possible life style of radiating embedded black holes during their continuous radiation processes.\\\\
{\bf Keywords:} Reissner-Nordstrom-Vaidya; Kerr-Newman-Vaidya black hole; rotating Vaidya; surface gravity; Hawking's radiation.
\end{abstract}


\section{Introduction}
In the theory of black hole expressed in spherical polar
coordinates, there is a singularity at the origin $r=0$; whereas
at $r=\infty$, the metric approaches the Minkowski flat
space. The event horizons of a black hole are expressed by this
coordinate. Thus, the nature of black holes depend on the radial
coordinate $r$. In an earlier paper [1] it is shown that Hawking's
radiation [2] can be expressed in classical space-time metrics, by
considering the charge $e$ to be function of the radial coordinate
$r$ of non-rotating Reissner-Nordstrom as well as rotating
Kerr-Newman black holes. The variable-charge $e(r)$ with respect
to the coordinate $r$ is followed from Boulware's suggestion [3]
that {\it the stress-energy tensor may be used to calculate the
change in the mass due to the radiation}. Accordingly,
the energy-momentum tensor of a particular space can
be used to calculate the change in the mass in order to
incorporate the Hawking's radiation effects in classical space-time
metrics. This idea suggests to consider the stress-energy tensor
of electromagnetic field of different forms or functions from
those of Reissner-Nordstrom, as well as Kerr-Newman, black holes
as these two black holes do not seem to have any direct Hawking's
radiation effects on the change of masses in space-time metrics. Thus,
a variable charge in the field equations
will have the different function of the energy-momentum tensor of
a charged black hole. {\sl Such a variable charge $e$ with
respect to the coordinate $r$ in Einstein's equations is referred
to as an electrical radiation ({\rm or} Hawking's electrical
radiation) of a black hole} [1]. For every electrical radiation
the charge $e$ is considered to be a function of $r$ in analyzing the
Einstein-Maxwell field equations and it has shown {\sl
mathematically} how the electrical radiation induces to produce
the changes of the masses of {\sl variable}-charged black holes.
The idea of losing (or changing) mass  can be incorporated in the
space-time metrics at the rate
as the electrical energy is radiated from the charged black hole.
In fact, the change in the mass of a charged black hole takes
place due to the vanishing of Ricci scalar of the electromagnetic
field. It was found that every electrical radiation $e(r)$ of the black hole leads
to a reduction in its mass by some quantity. If we consider such
electrical radiation taking place continuously for a long time,
then a continuous reduction of the mass will take place in the
black hole body, and the original mass of the black hole will
evaporate completely. At that stage the complete evaporation of
the mass will lead the gravity of the object depending only on the
electromagnetic field, and not on the mass of black hole.  Such an
object with zero mass  is being referred as an `instantaneous' naked singularity {\it i.e.} a naked singularity that exists for an instant and then continues
its electrical radiation to create negative mass [1]. So this
naked singularity is different from the one mentioned in
Steinmular {\it et al} [4], Tipler {\it et al} [5] in the sense
that an `instantaneous' naked singularity, discussed in [4, 5]
exists only for an instant and then disappears.

It is also noted that the time taken between two consecutive
radiations is supposed to be so short that one may not physically
realize how quickly radiations take place. Thus, it seems natural
to expect the existence of an `instantaneous' naked singularity
with zero mass only for an instant before continuing its next
radiation to create a negative mass naked singularity. This
suggests that it may also be possible in the common theory of
black holes that, as a black hole is invisible in nature, one may
not know whether, in the universe, a particular black hole has
mass or not, but electrical radiation may be detected on the black
hole surface. Immediately after the complete evaporation of the
mass, if one continues to radiate the remaining remnant, there may
be a formation of a new mass. If one repeats the electrical
radiation further, the new mass might increase gradually and then
the space-time geometry will represent the negative mass naked
singularity. The classical space-time metrics, for both stationary
rotating and non-rotating, which represent the negative mass naked
singularities have given in [1].

The black hole embedded into other spaces play an important role in relativistic framework. For example, the cosmological constant is found present in the inflationary scenario of the early universe in a stage where the universe is geometrically similar to the original de Sitter space [6]. Also black holes embedded into de Sitter cosmological universe can avoid the direct formation of negative mass naked singularities during Hawking's electrical radiation process of black holes [7]. It is also known that the rotating Vaidya black hole with variable mass is a non-stationary solution of Einstein's field equations and is non-vacuum {\it algebraically special} of Petrov classification of space-time [8, 9, 10]. In fact the non-stationary Vaidya black holes is commonly known with two different characters -- one is non-rotating [11] and other rotating [8, 9, 10]. This is due the fact that there are two stationary black holes vacuum solutions -- one is non-rotating Schwarzschild and other rotating Kerr. Hence it is highly necessary to understand the Vaidya black holes in two classes -- one is {\it non-rotating} and other {\it rotating}. This is because if we wish Vaidya black hole to embed into non-rotating Reissner-Nordstrom black hole, we need the non-rotating Vaidya black hole for similar configuration; whereas for rotating Kerr-Newman we should have rotating Vaidya black hole for the same rotating system.

The aim of this paper is to study Hawking's radiation of Reissner-Nordstrom and Kerr-Newman black hole embedded into the Vaidya backgrounds. The
solutions derived here describe non-rotating Reissner-Nordstrom black hole embedded into the non-rotating Vaidya to obtain a Reissner-Nordstrom-Vaidya  and  rotating embedded black hole
{\it i.e.}, the Kerr-Newman black hole is embedded into the rotating
Vaidya null radiating space generating the Kerr-Newman-Vaidya. The
definitions of these embedded black holes are in agreement with
the one defined by Cai {\it et.al.} [12], that {\sl when the
Schwarzschild black hole is embedded into the de Sitter space, one
has the Schwarzschild-de Sitter black hole}. Thus, these solutions
describe {\sl non-rotating}, Reissner-Nordstrom-Vaidya  and
rotating, Kerr-Newman-Vaidya  black holes.
These embedded solutions can be expressed in Kerr-Schild forms to
regard them as the extension of Glass-Krisch superposition [13], which is further
the extension of that of Xanthopoulos [14]. These Kerr-Schild
ansatze show that these embedded black holes are solutions of
Einstein's field equations. It is also noted that such generation
of embedded solutions in {\sl non-rotating} cases may be seen in
[15].

In an introductory survey Hawking and
Israel [16] have discussed the black hole radiation in three
possible ways with suggestive remarks --`So far there is no good
theoretical frame work with which to treat the final stages of a
black hole but there seem to be three possibilities: (i) The black
hole might disappear completely, leaving just the thermal
radiation that it emitted during its evaporation. (ii) It might
leave behind a non-radiating black hole of about the Planck mass.
(iii) The emission of energy might continue indefinitely creating
a negative mass naked singularity'. In order to examine
Hawking's radiation in classical space-time metrics, the
suggestion of Boulware [3] leads to consider the stress-energy tensors of electromagnetic field of different forms or functions from those of Reissner-Nordstrom, as well as Kerr-Newman, black holes as these two black holes do not seem to
have any direct Hawking's radiation effects. Considering the charge $e(r)$ to be a function of radial coordinate $r$, {\it i.e.} having different structure of the source of energy-momentum tensors, Hawking
electrical radiation of Reissner-Nordstrom and Kerr-Newman black holes has
been developed in the forms of space-time metrics in [1]. As a result, it appears that
(i) the changes in the mass of black holes, (ii) the formation of
`instantaneous' naked singularities  with zero mass and (iii) the
creation of `negative mass naked singularities' in
Reissner-Nordstrom as well as Kerr-Newman black holes may
presumably  be the correct formulations in classical space-time
metrics of the three possibilities of black hole evaporation
suggested by Hawking and Israel [16]. It has been  shown that, during the complete
evaporation of the masses of both Reissner-Nordstrom and Kerr-Newman black holes, the electrical radiation continues indefinitely creating negative mass naked singularities.
The same work of electrical radiation process has been extended in the presence of de Sitter cosmological background as Reissner-Nordstrom-de Sitter and Kerr-Newman-de Sitter in [7]. Due to the presence of the de Sitter background, the direct formation of `negative mass naked singularities' could be avoided by creating ``instantaneous'' charged de Sitter cosmological universe during radiation process. Here in this paper we are interested to develop a similar work of Hawking's electrical radiation on a different background having a variable mass function, rather than a constant on de Sitter background; {\it i.e.} Hawking's radiation of black holes on Vaidya background.

The paper is organized as follows:  In Sections 2 and 3, we develop embedded Reissner-Northrom-Vaidya and Kerr-Newman-Vaidya black holes by using Wang-Wu mass functions in Newman-Penrose spin coefficient formalism [17]. We discuss the physical properties of the solutions observing
the nature of their energy momentum tensors and Weyl scalars.
We also present the surface gravity, entropy and angular velocity
for each of these embedded black holes as they are important
parameters of a black hole. Sections 4 and  5 deal with the
Hawking's radiation on the variable-charged Reissner-Nordstrom-Vaidya and Kerr-Newman-Vaidya black holes.
We present various classical space-time metrics affected by the change
in the masses describing the possible life style of radiating
embedded black holes in different stages during radiation process.
In section 6 we conclude with a discussion of the results.
The relativistic aspect of Hawking's radiation has been studied in these embedded black holes.  In Appendix A, for a better explanation of the paper, we present an energy-momentum tensor for a metric having mass function $\hat{M}(u,r)$ of two variables. We also present the general equations for the energy conservation laws associated with $\hat{M}(u,r)$ in Newman-Penrose formalism. Appendix B is devoted for the NP quantities of the embedded Kerr-Newman-Vaidya solution and has shown that the energy-momentum tensors for the embedded black holes obey the energy conservation laws. The content of the paper are summarized in the form of theorems as:
\newtheorem{theorem}{Theorem}
\begin{theorem}
Embedded Reissner-Nordstrom-Vaidya is a Petrov type D in the classification of space-times, whose one of the repeated principal directions $\ell_a$ is geodesic, expanding, shear free and  zero twist.
\end{theorem}
\begin{theorem}
Embedded Kerr-Newman-Vaidya is a algebraically special type II in the Petrov classification of space-times, whose one of the repeated principal directions $\ell_a$ is geodesic,  expanding, shear free and non-zero twist.
\end{theorem}
\begin{theorem}
The horizon areas of the embedded Reissner-Nordstrom-Vaidya and Kerr-Newman-Vaidya are greater than the sum of the horizon areas of individual black holes, or the entropies of the embedded black holes are greater than the sum of the entropies of the individual ones.
\end{theorem}
\begin{theorem}
Embedded Reissner-Nordstrom-Vaidya and Kerr-Newman-Vaidya black holes can be expressed in Kerr-Schild ansatze in different backgrounds.
\end{theorem}
\begin{theorem}
Every electrical radiation of Reissner-Nordstrom-Vaidya  and Kerr-Newman-Vaidya  black holes will lead to a change in their masses without affecting the Maxwell scalar as well as Vaidya mass.
\end{theorem}
\begin{theorem}
The non-rotating and rotating charged Vaidya
metrics describe instantaneous naked singularities
during the Hawking's evaporation process of electrical
radiation of Reissner-Nordstrom-Vaidya and Kerr-Newman-Vaidya
black holes.
\end{theorem}
\begin{theorem}
During the radiation process,
after the complete evaporation of masses of both variable-charged
Reissner-Nordstrom-Vaidya and Kerr-Newman-Vaidya black
holes, the electrical radiation will continue indefinitely
creating negative mass naked singularities on Vaidya background.
\end{theorem}
\begin{theorem}
If an electrically radiating black hole,
rotating or non-rotating, is embedded into  Vaidya spaces, it
will continue to embed into the same space forever.
\end{theorem}
We find that Theorem 3 is consistent with the suggestion of Bekenstein [18] that ``when two black holes merge, the area of the resulting black hole (provided, of course, that one forms) cannot be smaller than the sum of the initial areas''. The idea of initial black holes comes from the embedded ones: in the case of embedded Reissner-Nordstrom-Vaidya, the Reissner-Nordstrom and the non-rotating Vaidya are regarded as initial (or individual) ones, and similarly Kerr-Newman and rotating Vaidya are the initial (or individual) ones in Kerr-Newman-Vaidya.
Theorem 4 shows that embedded Reissner-Nordstrom-Vaidya and Kerr-Newman-Vaidya black holes are solutions of Einstein's field equations. It is found that the Theorems 5, 6 and 7 are in favour of the three possibilities of black holes made by Hawking and Israel [16], and indicate the various
stages of the life of embedded radiating black
holes. Theorem 7 provides a violation of Penrose's cosmic-censorship hypothesis that``no naked singularity can ever be created"[19]. Theorem 8 shows the nature of embedded radiating black holes that continue to embed into the same space.
Here it is convenient to use the phrase `change in
the mass' rather than `loss of mass' as there may be a
possibility of creation of mass after the exhaustion of the
original mass if one continues the same process of electrical
radiation. This will be seen later in the paper. The presentation
of this paper is essentially based on the differential form technique developed by McIntosh and Hickman [20] in $(-2)$ signature.

\section{Reissner-Nordstrom-Vaidya Solution}
\setcounter{equation}{0}
\renewcommand{\theequation}{2.\arabic{equation}}
\vspace*{.15in}
In this section for deriving an embedded Reissner-Nordstrom-Vaidya we consider a general canonical metric in Eddington-Finkestein coordinate system given in (A.2) of Appendix A, with the quantities $\mu^*, \rho^*$ and $p$ (when the rotational parameter $a$ sets to zero) in (A.5). However, there is no straightforward way to derive an embedded Reissner-Nordstrom-Vaidya solution by solving the non-linear Einstein's field equations associated with the line element (A.2) having a variable mass function $M(u,r)$. Therefore, in order to embed the Reissner-Nordstrom into Vaidya space, we choose, without loss of generality, the $q_n(u)$ in the Wang-Wu mass function $M(u,r)$ [15] as follows
\begin{eqnarray}
\begin{array}{cc}
q_n(u)=&\left\{\begin{array}{ll}
M+f(u),&{\rm when }\;\;n=0\\
-e^2/2, &{\rm when }\;\;n=-1\\
0, &{\rm when }\;\;n\neq 0, -1,
\end{array}\right.
\end{array}
\end{eqnarray}
where $M$ and $e$ are constants. Then, the  mass function takes the form
\begin{equation}
\hat{M}(u,r)=\sum_{n=-\infty}^{+\infty}
q_n(u)\,r^n =M+f(u)-\frac{e^2}{2r}.
\end{equation}
Using this mass function in general canonical metric (A.2) in Eddington-Finkestein coordinate system, we find a line element
\begin{eqnarray}
d s^2&=&[1-r^{-2}\{2r(M+f(u))-e^2\}]\,du^2 +2du\,dr \cr
&&-r^2d\theta^2 -r^2{\rm sin}^2\theta\,d\phi^2.
\end{eqnarray}
The constants $ M$ and $e$ are related with the mass and charge of Reissner-Nordstrom solution respectively.  $f(u)$ is the variable mass of Vaidya solution, $u$ is the retarded time which is used for calculating the outgoing or radiating energy-momentum tensor. When $f(u)$ sets to zero, the above line element will reduce to the Reissner-Nordstrom solution. The choice of mass function in (2.2) is acceptable in the framework of general relativity as the linear superposition of particular known solutions in $M(u,r)$ is also a solution of Einstein's field equations.

Now by using (2.) in (A.3-A.5) when $a=0$, we find the energy-momentum tensor for the Reissner-Nordstrom-Vaidya solution as
\begin{eqnarray}
T_{ab}=\mu^*\ell_a\ell_b + 2\,\rho^*\ell_{(a}n_{b)}+ 2 p\, m_{(a}\bar{m}_{b)}
\end{eqnarray}
which can be expressed in two parts as
\begin{eqnarray}
T_{ab}^{(\rm n)}&=&\mu^*\ell_a\ell_b,\\
T_{ab}^{(\rm E)}&=&2\,\rho^*\ell_{(a}n_{b)}+2 p\, m_{(a}\bar{m}_{b)},
\end{eqnarray}
where the null density $\mu^*$, the energy density $\rho^*$ and pressure $p$ are found as follows
\begin{eqnarray}
&&\mu^*=-\frac{2}{K\,r^2}f(u)_{,u}, \quad \rho^{*}=p=\frac{e^2}{K\,r^4}.
\end{eqnarray}
Here $\ell_a$ and $n_a$ are real, and $m_a$ is complex null vectors with the normalization conditions $\ell_an^a=1=-m_a\bar m^a$; and they are given by
\begin{eqnarray}
&&\ell_a=\delta^1_a ,\;n_a=\Big[1-r^{-2}\Big\{2r(M+f(u))-e^2\Big\}\Big]\,\delta^1_a+ \delta^2_a, \cr
&&m_a=-{r\over\surd 2}\,\Big\{\delta^3_a +i\,{\rm
sin}\,\theta\,\delta^4_a\Big\}.
\end{eqnarray}
We use the notation $2\ell_{(a}n_{b)}=\ell_a n_b+n_a \ell_b$.  $T_{ab}^{(\rm n)}$ is the energy-momentum tensor for the Vaidya null radiating fluid and $T_{ab}^{(\rm E)}$  for non-null electromagnetic field existing in non-rotating Reissner-Nordstrom-Vaidya metric $g_{ab}$. The energy-momentum tensor (2.4) can be regarded as the electromagnetic field $T_{ab}^{(\rm E)}$ is interacting with the null radiating fluid $T_{ab}^{(\rm n)}$, and is traceless $(T=0)$. $T^{ab}$ in (2.4)
obeys the energy conservation equations
\begin{equation}
T^{ab}_{\;\;\;\,;b}=0.
\end{equation}
showing that the non-rotating Reissner-Nordstrom-Vaidya metric is an exact solution of Einstein's field equation.

We also fine the Weyl scalar for the solution (2.3) as follows
\begin{eqnarray}
\psi_2=-\frac{1}{r^4}\{r(M+f(u))-e^2\}.
\end{eqnarray}
The non-vanishing of Weyl scalar $\psi_2$ means that the Reissner-Nordstrom-Vaidya solution is Petrov type D whose one of the repeated principal directions $\ell_a$ is geodesic $(\kappa=\epsilon=0)$, shear free $(\sigma=0)$ and  expanding $(\hat{\theta}\neq 0)$ with zero twist $(\hat{\omega}=0)$. (Here $\kappa, \epsilon, \sigma,$ etc. are Newman-Penrose spin coefficients for the line element (2.3) and can be obtained from (B.2) by setting rotational parameter $a=0$.) This completes the proof of the Theorem 1.

The invariants of Ricci and Reimann Curvature tensors for the solution are given as follows
\begin{subequations}
\label{eq:whole}
\begin{eqnarray}
&&R_{ab}R^{ab}=\frac{4}{r^8}e^4 \\ \label{subeq:1}
&&R_{abcd}R^{abcd}={1\over r^8}\Big[48\Big(r\{M+f(u)\}-e^2\Big)^2+ 20 e^4\Big]. \label{subeq:2}
\end{eqnarray}
\end{subequations}
These invariants becomes infinite when $r\rightarrow 0$. $K \rightarrow \infty$, which  suggests that the Reissner-Nordstrom-Vaidya is a model of a primordial black hole [21].

The solution of Reissner-Nordstrom-Vaidya will describe a black hole with external horizon $r_+$ and internal Cauchy horizon $r_-$ given by
\begin{equation}
r_{\pm}=\{M+f(u)\}\pm\surd {[\{M+f(u)\}^2-e^2]},
\end{equation}
when $\Delta\equiv r^2-2r\{M+f(u)\}-e^2=0$.
The areas at these horizons are obtained as
\begin{eqnarray}
A^{\rm RNV}_{\pm}&=&\int^\pi_0 \int^{2\pi}_0\sqrt{g_{\theta\theta}g_{\phi\phi}}\,\, d\theta d\phi\Big|_{r=r_{\pm}} \cr
&=&8\pi\Big\{(M+f(u))^2-\frac{e^2}{2}\pm(M+f(u))\sqrt{(M+f(u))^2-e^2}\Big\}.
\end{eqnarray}
It is observed that when $f(u)$ is set to zero, the area $A^{\rm RNV}_{\pm}$ will reduce to that of Reissner-Nordstrom black hole, whereas if we set $M$ and $e$ to be zero, the area $A^{\rm RNV}_{\pm}$ (2.13) will be that of Vaidya black hole as
\begin{subequations}
\label{eq:whole}
\begin{eqnarray}
&&A^{\rm RN}_{\pm}= 8 \pi\{M^2-\frac{e^2}{2}\pm M\sqrt{M^2-e^2}\},\\ \label{subeq:1}
&&A^{\rm V}_{+}= 16 \pi f^2(u)\quad{\rm and } \quad A^{\rm V}_{-}=0. \label{subeq:2}
\end{eqnarray}
\end{subequations}
From (2.13) and (2.14) we find that the area $A^{\rm RNV}_{\pm}$ of the embedded Reissner-Nordstrom-Vaidya black hole is greater than the sum of the areas of the individual ones {\it i.e}
\begin{equation}
A^{\rm RNV}_{\pm}>A^{\rm RN}_{\pm}+A^{\rm V}_{\pm}.
\end{equation}
which proves Theorem 3 for non-rotating case stated above in the introduction. This greater area theorem of the embedded black hole is in agreement with Bekenstein's suggestion [18]. Now
from the entropy-area relation $S=A/4$ [22], we obtain the entropies of the horizons
\begin{eqnarray}
S^{\rm RNV}_\pm&=&2 \pi\big[\{M+f(u)\}r_{\pm}-{e^2 \over 2}\big],
\end{eqnarray}
which satisfy the equality relations $S^{\rm RNV}_{\pm}>S^{\rm RN}_{\pm}+S^{\rm V}_{\pm}$. Here $S^{\rm RN}_{\pm}$ and $S^{\rm V}_{\pm}$ denote the entropies of Reissner-Nordstrom and Vaidya black holes respectively.

The surface gravity ${\cal \kappa}$ of a horizon is defined by the relation $n^b\nabla_b n^a={\cal \kappa}n^a$, where the null vector $n^a$ in (2.8) above is parameterized by the coordinate $u$, such that $d/du=n^b\nabla_b$ [23, 24].
The surface gravities of the horizons at $r=r_{+}$ and $r=r_{-}$ are found as
\begin{eqnarray}
{\cal K}_{+}=-\frac{1}{2 r_{+}^2}(r_{+} - r_{-}),
\quad
{\cal K}_{-}=\frac{1}{2 r_{-}^2}(r_{+} - r_{-})
\end{eqnarray}
and their ratio is
\begin{equation}
{\cal K}=-\frac{r^2_{+}}{r^2_{-}}
\end{equation}
Hawking's temperature at horizons  $r_{+}$ and $r_{-}$ are given by
\begin{eqnarray}
T_{+}=\frac {{\cal K}_{+}}{2 \pi}=-\frac {1}{4 \pi r_{+}^2}(r_{+}-r_{-}), \quad
T_{-}=\frac {{\cal K}_{-}}{2 \pi}=\frac {1}{4 \pi r_{-}^2}(r_{+}-r_{-}).
\end{eqnarray}
Since there are two horizons, there must be two different temperatures given by $T_{+}$ and $T_{-}$ from the event and Cauchy horizon respectively. It is also observed that the ratio of the temperatures at the horizons is same as the ratio of the surface gravities.

The non-rotating Reissner-Nordstrom-Vaidya solution can be expressed in
Kerr-Schild anstaz on the Reissner-Nordstrom background $g^{RN}_{ab}$ as
\begin{equation}
g_{ab}^{\rm RNV}=g_{ab}^{\rm RN} -\{2f(u)/r\}\ell_a\ell_b.
\end{equation}
Also the metric can be expressed on Vaidya background $g^V_{ab}$ as
\begin{equation}
g_{ab}^{\rm RNV}=g_{ab}^{\rm V} -\{2M/r-e^2/r^2\}\ell_a\ell_b.
\end{equation}
The expressibility of these ansatze on different background spaces, means that it is always true to talk about either Reissner-Nordstrom black hole embedded into the Vaidya space as Reissner-Nordstrom-Vaidya or the Vaidya black hole embedded into the Reissner-Nordstrom space to form Vaidya-Reissner-Nordstrom -- both possess the same geometrical meaning. This indicates that in the nature of the embedded black hole, one cannot physically predict which black hole starts first to embed into another. It is also found that the two metric tensors of Reissner-Nordstrom and Vaidya solutions cannot be added to emerge into an embedded Reissner-Nordstrom-Vaidya solution as
\begin{equation}
g_{ab}^{\rm RNV}\neq{1\over2}\Big(g_{ab}^{\rm RN}+g_{ab}^{\rm V}\Big);\nonumber
\end{equation}
however the energy-momentum tensors of the respective solutions can be added to obtain that of the  embedded one. It is to mention that the Kerr-Schild ansatze (2.20) and (2.21) are the generalizations of
Xanthopoulos [14] as well as Glass and Krisch [13] in the charged
Reissner-Nordstrom solution.
These ansatze shows the proof of the non-rotating part of Theorem 2 above.

\section{Kerr-Newman-Vaidya Solution}
\setcounter{equation}{0}
\renewcommand{\theequation}{3.\arabic{equation}}
In this section, we shall develop an embedded Kerr-Newman-Vaidya black hole which will be the generalization of the non-rotating Reissner-Nordstrom-Vaidya solution (2.3), by using the rotating metric (A.1) of Appendix A with a mass function $M(u,r)$ and the quantities $\mu^*$, $\rho^*$. $p$ and $\omega^*$ in (A.5). However, there is no straightforward way of solving the Einstein's field equations associated with the line element (A.1) for obtaining a Kerr-Newman-Vaidya solution. Hence, in order to obtain the rotating solution, we utilize the mass function (2.2) in the rotating frame (A.1) given in Appendix A and find a line element describing Kerr-Newman-Vaidya solution as
\begin{eqnarray}
d s^2&=&[1-R^{-2}\{2r(M+f(u))-e^2\}]\,du^2 +2du\,dr
+2aR^{-2}\{2r(M+f(u)) \cr &&-e^2\}\,{\rm sin}^2 \theta\,du\,d\phi
-2a\,{\rm sin}^2\theta\,dr\,d\phi -R^2d\theta^2 \cr
&&-\{(r^2+a^2)^2 -\Delta a^2\,{\rm sin}^2\theta\}\,R^{-2}{\rm
sin}^2\theta\,d\phi^2,
\end{eqnarray}
where $\Delta=r^2-2r\{M+f(u)\}+a^2+e^2$.
Here $M$ and $e$ are the mass and the charge of Kerr-Newman solution, $a$ is the
rotational parameter  and $f(u)$ represents the mass
function of rotating Vaidya null radiating fluid. When we set $f(u)=0$, the metric (3.1) recovers the usual Kerr-Newman black hole, and if $M=0$, then it is the `rotating' charged Vaidya null radiating black hole [8, 9, 10]. The line element (3.1) describes the embedded Kerr-Newman-Vaidya black hole having singularity when $\Delta=0$.

In this rotating solution, the Vaidya null fluid is interacting with the non-null electromagnetic field $F_{ab}$, whose Maxwell scalar is defined by $\phi_1={1\over 2}F_{ab}(\ell^a\,n^b+\overline{m}^a\,m^b)$. Then, by using (2.2) in (A.3) we find an  energy-momentum tensor for the rotating line element (3.1) as follows:
\begin{eqnarray}
T_{ab}&=&\mu^*\ell_a\ell_b + 2\omega\ell_{(a}\bar{m}_{b)}+2\bar{\omega}\ell_{(a}m_{b)}
+2\rho^*\ell_{(a}\,n_{b)} + 2 p\, m_{(a}\,\bar{m}_{b)}.
\end{eqnarray}
Here $\ell_a$ and $n_a$ are real and $m_a$ is complex null vectors with the normalization conditions $\ell_an^a=1=-m_a\bar m^a$, and are given in (B.1).
This $T_{ab}$ may be expressed in terms of those
of the electromagnetic field as well as the rotating Vaidya null fluid
respectively as
\begin{eqnarray}
&&T^{(\rm E)}_{ab}=2\rho^*\ell_{(a}\,n_{b)} + 2p\,m_{(a}\,\bar{m}_{b)}\\
&&T^{(\rm n)}_{ab}= \mu^*\ell_a\ell_b + 2\omega\ell_{(a}\bar{m}_{b)}+2\bar{\omega}\ell_{(a}m_{b)},
\end{eqnarray}
where the quantities are
\begin{subequations}
\label{eq:whole}
\begin{eqnarray}
&&\rho^*=p={e^2\over K\,R^2\,R^2}, \\ \label{subeq:1}
&&\mu^*=-\frac{r}{K\,R^2\,R^2}\Big\{2\,r\,f(u)_{,u}+a^2{\rm
sin}^2\theta\,f(u)_{,uu}\Big\},\\ \label{subeq:2}
&&\omega =-\frac{i\,a\,
\sin\theta}{\surd 2\,K\,\bar R\,R^2}\,f(u)_{,u}.\label{subeq:3}
\end{eqnarray}
\end{subequations}
Here $\rho^*$ and $p$ are denoted respectively the energy density and pressure of the electromagnetic field of Kerr-Newman black hole, and $\mu^*$ being the null density contributed from the Vaidya mass $f(u)$. $\omega$ is the rotational density arisen from the parameter $a$, that the null
fluid $T^{(\rm n)}_{ab}$ is rotating as the expression of $\omega$ involves the rotating parameter $a$ coupling with $\partial f(u)/\partial u$; both are non-zero quantities
for the existence of the rotating Vaidya null radiating solution. It is to mention that
the energy-momentum tensor (3.2) obeys the energy conservation equations
\begin{equation}
T^{ab}_{\;\;\;\,;b}=0.
\end{equation}
as shown in (B.7) of Appendix B. This indicates that the metric tensor of Kerr-Newman-Vaidya associated with the line element (3.1) is a solution of Einstein's field equations.

When $\Delta=0$, the solution (3.1) will describe an embedded black hole
with external event horizon $r_+$ and internal Cauchy horizon $r_-$ which are found as
\begin{equation}
r_{\pm}= \{M+f(u)\}\pm\surd {[\{M+f(u)\}^2-a^2-e^2]},
\end{equation}
and the non-stationary limit surface $g_{uu}=0$ is given by
\begin{equation}
r\equiv r_e(u,\theta)= \{M+f(u)\}
+\surd {[\{M+f(u)\}^2-a^2{\rm cos}^2\theta-e^2]}.
\end{equation}
which coincides with the event horizon $r_+$ only at the points $\theta=0$ and
$\theta=\pi$. This feature distinguishes the Kerr-Newman-Vaidya space-time from the Reissner-Nordstrom-Vaidya one. This situation cannot be observed in non-rotating solution. The area of the horizon is  found as
\begin{eqnarray}
A^{\rm KNV}_{\pm}&=&\int^\pi_0 \int^{2\pi}_0\sqrt{g_{\theta\theta}g_{\phi\phi}} d\theta d\phi\Big|_{r=r_{\pm}} \cr
&=& 4\pi\Big\{(M+f(u))^2-e^2\pm 2(M+f(u))\sqrt{(M+f(u))^2-a^2-e^2}\Big\}.  \nonumber \\ \end{eqnarray}
When $f(u)$ is set to zero, the area $A^{\rm KNV}_{\pm}$ will reduce to that of Kerr-Newman black hole $A^{\rm KN}_{\pm}$; whereas if we set $M$ and $e$ to be zero the area will be that of the rotating Vaidya black holes $A^{\rm V}_{\pm}$ as
\begin{subequations}
\label{eq:whole}
\begin{eqnarray}
&&A^{\rm KN}_{\pm}=4 \pi\Big\{2M^2-e^2\pm 2M\sqrt{M^2-a^2-e^2}\Big\}\\ \label{subeq:1}
&&A^{\rm V}_{\pm}= 4 \pi\Big\{2f^2(u)\pm 2f(u)\sqrt{f^2(u)-a^2}\Big\}. \label{subeq:2}
\end{eqnarray}
\end{subequations}
We observe the difference between the areas of the rotating (3.9) and non-rotating (2.14b) Vaidya solutions, that the non-rotating Vaidya does not have $A^{\rm V}_{-}$.
From (3.9) and (3.10), we find that the area of embedded black hole is greater than the sum of the areas of the individual black holes as
\begin{equation}
A^{\rm KNV}_{\pm}> A^{\rm KN}_{\pm}+A^{\rm V}_{\pm}
\end{equation}
which establishes the proof of Theorem 3 for the rotating part.
The inequality relation (3.11) is consistent with Bekenstein's suggestion [18] that, {\it if two black holes emerges to form a black hole, the area of the later will be greater than the sum of the individual ones}.
Now, from the entropy-area relation of Bekenstein and Hawking [22], we obtain the entropy
\begin{eqnarray}
S^{\rm KNV}_\pm=2\pi\Big[\{M+f(u)\}r_{\pm}-\frac{e^2}{2}\Big].
\end{eqnarray}
This entropy satisfies the inequality relation $S^{\rm KNV}_{\pm}> S^{\rm KN}_{\pm}+S^{\rm V}_{\pm}$, which is true for the rotating and non-rotating embedded black holes. It is also found that the sum of the entropies at the event ($r_+$) and Cauchy ($r_-$) horizons does not affect by the rotational parameter $a$ as $S=S^{\rm KNV}_{r_+}+S^{\rm KNV}_{r_-}=4\pi{(M+f(u))^2}-2\pi e^2$.

The surface gravity of the horizons at $r=r_{\pm}$ are
\begin{eqnarray}
{\cal K}_{+}=\frac{1}{2 R_{+}^2}(r_{-} - r_{+}),
\quad
{\cal K}_{-}=\frac{1}{2 R_{-}^2}(r_{+} - r_{-})
\end{eqnarray}
where $R_{\pm}^2= r_{\pm}^2+a^2\cos^2\theta$.
The ratio of the two surface gravities, ${\cal K}$ is
\begin{eqnarray}
{\cal K}&=&\frac{{\cal K}_{+}}{{\cal K}_{-}}=-\frac {R_{-}^2}{R_{+}^2}.
\end{eqnarray}
According to  Padmanabhan  and Roy Choudhury [25], there is a
global temperature if the ratio is rational.
Hawking's temperatures associated with ${\cal K}_{+}$, and ${\cal K}_{-}$ are obtained as
\begin{eqnarray}
T_{+}=\frac {{\cal K}_{+}}{2 \pi} =\frac {r_{-}-r_{+}}{4 \pi R_{+}^2}, \quad
T_{-}=\frac {{\cal K}_{-}}{2 \pi} =-\frac {r_{-}-r_{+}}{4 \pi R_{-}^2}.
\end{eqnarray}
This shows that there are two different temperatures -- one flowing from the event horizon while other  from the Cauchy horizon. It is also observed that the ratio of the temperatures is equal to the ratio of the surface gravities (3.14).
The angular velocity of the horizons take the form
\begin{eqnarray}
\Omega_{\pm}=\lim_{r \to r_{\pm}}\Big(-\frac{g_{u\phi}}{g_{\phi\phi}}\Big)
=\frac{a[2r\{M+f(u)\}-e^2]}{(r^2+a^2)^2}\Big|_{r=r_{\pm}}.
\end{eqnarray}
This will vanish when the rotational parameter $a$ sets to zero. This shows the fact that the non-rotating solution does not have any angular velocity. For instance, the Reissner-Nordstrom-Vaidya black hole discussed above has no angular velocity.

The rotating Kerr-Newman-Vaidya metric (3.1) can be
expressed in Kerr-Schild form on the Kerr-Newman background as
\begin{equation}
g_{ab}^{\rm KNV}=g_{ab}^{\rm KN} +2Q(u,r,\theta)\ell_a\ell_b
\end{equation}
where $Q(u,r,\theta)=-rf(u)R^{-2}$, and the vector $\ell_a$ is a geodesic, shear free,
expanding as well as rotating null vector of both
$g_{ab}^{\rm KN}$ as well as $g_{ab}^{\rm KNV}$.
$g_{ab}^{\rm KN}$ is the Kerr-Newman metric
with $m=e$= constant. This null vector $\ell_a$ is
one of the double repeated principal
null vectors of the Weyl tensor of $g_{ab}^{\rm KN}$.
It appears that the rotating Kerr-Newman geometry may be regarded
as joining smoothly with the rotating Vaidya geometry at its null
radiative boundary, as shown by Glass and Krisch [13] in the case
of Schwarzschild geometry joining to the non-rotating Vaidya
space-time. The Kerr-Schild form (3.17) will recover that of
Xanthopoulos [14] $g'_{ab}=g_{ab}+\ell_a\ell_b$, when
$Q(u,r,\theta) \rightarrow 1/2$ and that of Glass and Krisch [13]
$g'_{ab}=g_{ab}^{\rm Sch}-\{2f(u)/r\}\ell_a\ell_b$ when $e=a=0$
for non-rotating Schwarzschild background space. Thus, one can
consider the Kerr-Schild form (3.17) as the extension of those of
Xanthopoulos as well as Glass and Krisch. When we set $a=0$, this
rotating Kerr-Newman-Vaidya solution (3.1) will recover to
non-rotating Reissner-Nordstrom-Vaidya solution (2.3) with the
Kerr-Schild form $g_{ab}^{\rm RNV}=g_{ab}^{\rm RN} -
\{2f(u)/r\}\ell_a\ell_b$, which is still a generalization of
Xanthopoulos as well as Glass and Krisch in the charged
Reissner-Nordstrom solution.

To interpret the Kerr-Newman-Vaidya solution as the Kerr-Newman black hole
embedded into the rotating Vaidya background, we can write
the Kerr-Schild form (3.17) as
\begin{equation}
g_{ab}^{\rm KNV}=g_{ab}^{\rm V} +2Q(r,\theta)\ell_a\ell_b
\end{equation}
where $Q(r,\theta)=-(rM - e^2/2)R^{-2}$.
Here,  $g_{ab}^{\rm V}$ is the rotating
Vaidya null radiating black hole and $\ell_a$ is the geodesic null vector for both $g_{ab}^{\rm KNV}$ and $g_{ab}^{\rm V}$ {\it.i.e.} $g_{ab}^{\rm KNV}\ell^a\ell^b=0=g_{ab}^{\rm V}\ell^a\ell^b$.
When we set $f(u)=a=0$, $g_{ab}^{\rm V}$ will recover the flat
metric, then  $g_{ab}^{\rm KNV}$ becomes the original Kerr-Schild
form written in spherical symmetric flat background. It is worth to mention the fact that the rotating embedded
solution (3.1) cannot be considered as a bimetric theory since
\begin{equation}
g_{ab}^{\rm KNV}\neq{1\over2}\Big(g_{ab}^{\rm KN}+g_{ab}^{\rm V}\Big).
\end{equation}
This indicates the fact that one cannot simply add the two solutions $g_{ab}^{\rm KN}$ and $g_{ab}^{\rm V}$ in order to get embedded solution $g_{ab}^{\rm KNV}$. However their energy momentum tensors $T^{(n)}_{ab}$ and $T^{(\rm E)}_{ab}$ given in (3.3) and (3.4) respectively can be added to obtain $T_{ab}$ as in (3.2) above.

These two Kerr-Schild forms (3.17) and (3.18) certainly confirm
that the metric  $g_{ab}^{\rm KNV}$  is a solution of Einstein's
field equations since the background rotating metrics $g_{ab}^{\rm
KN}$ and  $g_{ab}^{\rm V}$ are solutions of Einstein's equations.
They both possess different stress-energy tensors $T_{ab}^{(\rm
n)}$ and $T_{ab}^{(\rm E)}$ given in (3.3) and (3.4)
respectively. Looking at the Kerr-Schild form (3.18), the
Kerr-Newman-Vaidya black hole can be treated as a generalization
of Kerr-Newman black hole by incorporating Visser's suggestion
[26] that {\it Kerr-Newman black hole embedded in an axisymmetric
cloud of matter would be of interest}. The expressibility of this embedded black hole in different ansatze (3.17), (3.18) means that, it is always true to talk about either Kerr-Newman black hole embedded into the Vaidya space as Kerr-Newman-Vaidya or the Vaidya black hole embedded into the Kerr-Newman space to form Vaidya-Kerr-Newman -- both  nomenclature possess the same geometrical meaning. That is, one cannot physically predict which black hole started first to embed into another black hole. These Kerr-Schild ansatze (3.17) and (3.18) establish the proof of the rotating part of the Theorem 2.

The line element of Kerr-Newman-Vaidya black hole can also be expressed in the coordinate system $(t, x, y, z)$ as
\begin{eqnarray}
ds^2&=&dt^2-dx^2-dy^2-dz^2-\frac{2r^3}{r^4+a^2z^2}\Big\{M+f(u)-\frac{e^2}{2r}\Big\}\\&&\times \Big[dt-\frac{1}
{r^2+a^2}\Big\{r(x dx+ y dy)+ a(x dy-y dx)\Big\}-\frac{z}{r}dz\Big]^2 \nonumber
\end{eqnarray}
where $r$ is defined in terms of $x$, $y$ and $z$ as
\begin{equation}
r^4-(x^2+ y^2+ z^2- a^2)r^2- a^2z^2=0 \nonumber
\end{equation}
with the following transformations
\begin{eqnarray}
&&x=(r\cos\phi+a\sin\phi)\sin\theta,\cr
&&y=(r\sin\phi-a\cos\phi)\sin\theta,\\
&&z=r\cos\theta, \quad t=u+r. \nonumber
\end{eqnarray}
Then, the above transformed metric (3.20) can be written in the Kerr-Schild form on flat background as
\begin{equation}
g^{\rm KNV}_{ab}=\eta_{ab}+2H(t, x, y,z)\ell_a \ell_b,
\end{equation}
where $\eta_{ab}$ is the flat metric and
\begin{eqnarray}
&&H(t, x, y, z)= -\frac{2r^3}{r^4+a^2z^2}\Big\{M+f(t,r)-\frac{e^2}{2r}\Big\},\cr
&&\ell_a {dx}^a=dt-\frac{1}{(r^2+a^2)}\Big\{r(x dx+y dy)+a(x dy-y dx)\Big\}-\frac{1}{r}z dz.
\end{eqnarray}
Here $\ell_a$ is the null vector with respect to $g^{\rm KNV}_{ab}$ and $\eta_{ab}$, and is given in null coordinate system $(u, r, \theta, \phi)$ in  Appendix A below.
The Kerr-Schild form (3.22) shows that the Kerr-Newman-Vaidya solution expressed in coordinate system (3.21) is a solution of Einstein's field equations. The transformed metric (3.20) is analytic everywhere except at $x^2+y^2+z^2=a^2$ and $z=0$.

\section{Changing mass of variable charged \\Reissner-Nordstrom-Vaidya black hole}
\setcounter{equation}{0}
\renewcommand{\theequation}{4.\arabic{equation}}
In this section, as a part of discussion of the physical
properties of the embedded solutions, we
shall discuss a scenario which is capable of avoiding the
formation of negative mass naked singularity during Hawking
radiation process in Reissner-Nordstrom-Vaidya space-time metrics, describing the life style
of embedded radiating black hole. The formation of
negative mass naked singularities in classical space-time metrics
are being shown in [1] after the complete evaporation of the masses
of {\sl non-embedded} rotating Kerr-Newman and non-rotating
Reissner-Nordstrom, black holes due to Hawking electrical
radiation.

Here we shall clarify two similar nomenclatures having different
meaning like Hawking's radiation and Vaidya null radiation.
Hawking's radiation is continuous radiation process of energy from the
black hole body thereby leading to the change in mass [1-5],
whereas the Vaidya null radiation means that the stress-energy
momentum tensor describing the gravitation in Vaidya space-time
metric is a null radiating fluid [11]. So Vaidya null radiation
does not have any direct relation with Hawking's radiation of
black holes [18].
By electrical radiation of a charged black hole we mean the
variation of the charge $e$ with respect to the coordinate $r$
in the stress-energy momentum tensor of electromagnetic field, thereby able to establish the change of mass of the black hole.

As mentioned earlier in the introduction,
the variation of the charge $e$ will certainly lead different forms or functions of
the stress-energy tensor from that of Reissner-Nordstrom-Vaidya black
hole. To observe the change in the mass of black hole in
the space-time metric, one has to consider a different form or
function of stress-energy tensor of a particular black hole. That
is, in order to incorporate the Hawking's radiation in
this black hole (2.3), we must have a different stress-energy
tensor as the Reissner-Nordstrom-Vaidya black hole with $T_{ab}$
(2.4) does not show any direct Hawking's radiation effects.
The consideration of different forms of stress-energy-momentum
tensor in the study of Hawking's radiation effect in classical
space-time metrics here is in agreement with Boulware's suggestion
[3] mentioned in introduction above.

The line element of the variable-charged Reissner-Nordstrom-Vaidya
solution with the assumption that the charged $e$ is a function of coordinate r in (2.3),
is given by
\begin{eqnarray}
d s^2&=&[1-r^{-2}\{2r(M+f(u))-e^2(r)\}]\,du^2 +2du\,dr \cr
&&-r^2d\theta^2 -r^2 \,{\rm sin}^2\theta\,d\phi^2.
\end{eqnarray}
If we set $e(r)$ = constant initially, the metric will recover the charged
Reissner-Nordstrom-Vaidya solution (2.3). Here with the variable charge $e(r)$ in (4.1) we find the different  energy-momentum tensor from that of the Reissner-Nordstrom-Vaidya solution. Accordingly, from the Cartan's second equations, we calculate the tetrad components of Ricci and Weyl tensors and found  as follows:
\begin{eqnarray}
\phi_{11}&=&{1\over
{2\,R^2\,R^2}}\,\Big\{e^2(r)-2r\,e(r)e(r)_{,r}\Big\}
+{1\over 4R^2}\,\Big\{e^2(r)_{,r}+e(r)\,e(r)_{,rr}\Big\} \cr
\phi_{22}&=&-\frac{f(u)_{,u}}{r^2} \\\
\Lambda&=& -{1\over
12\,r^2}\,\Big\{e^2(r)_{,r}+e(r)\,e(r)_{,rr}\Big\},\\\
\psi_2&=&{1\over\bar{R}\,\bar{R}\,R^2}\Big[e^2(r)-R\{M+f(u)\}+{1\over 6}\,\bar{R}\,\bar{R}\,\Big\{e^2(r)_{,r}
+e(r)\,e(r)_{,rr}\Big\} \cr &&-\bar{ R}\,e(r)\,e(r)_{,r}\Big].
\end{eqnarray}
Here we find the non-zero Ricci scalar $\Lambda=\frac{1}{24}g^{ab}R_{ab}$ in (4.3). However, for an electromagnetic field, this Ricci scalar $\Lambda$ has to vanish leading to the equation
\begin{equation}
e^2(r)_{,r}+e(r) e(r)_{,rr}=0
\end{equation}
which can be solved having the solution
\begin{equation}
e^2(r)=2rm_1 + C
\end{equation}
where $m_1$ and $C$ are real constants. Substituting this
$e^2(r)$ in (4.2) and (4.4), and after identifying the
constant $C\equiv e^2$, we obtain
\begin{eqnarray}
&&\phi_{11}=\frac{1}{2\,r^4}{e^2}, \\
&&\psi_2=-{1\over{r^4}}\Big[r\{M-m_1+f(u)\}-e^2\Big].
\end{eqnarray}
and $\phi_{22}$ is remained
unaffected as in (4.2) above. Accordingly,  from (4.7) the Maxwell
scalar $\phi_1$ with constant $e$ becomes
\begin{eqnarray}
\phi_{1}={e\over {\surd 2\,r^2}}
\end{eqnarray}
since $\phi_{11}=\phi_{1}\bar{\phi_1}$ for electromagnetic field.
However, from the Weyl scalar (4.8) we have clearly seen a change in the mass M by some constant quantity $m_1$  for the first electrical radiation in the embedded black hole. Then the total
mass of the  Reissner-Nordstrom-Vaidya black hole becomes $(M-m_{1})+f(u)$ and the metric takes the form
\begin{eqnarray}
d s^2&=&[1-r^{-2}\{2r(M-m_1+f(u))-e^2\}]\,du^2 +2du\,dr \cr
&&-r^2d\theta^2 -r^2 \,{\rm sin}^2\theta\,d\phi^2.
\end{eqnarray}
Since, for the first radiation, the Maxwell scalar (4.9) remain the same,
radiation, we again consider the charge $e$ to be a function of $r$ for the second in the above metric (4.10) which certainly leads, by  the Einstein-Maxwell field equations with the vanishing of $\Lambda$, to reduce another quantity $m_2$ (say) from the total mass, {\it i.e.} after the second radiation the mass becomes $M-m_{1}-m_{2}+f(u)$. Again, the Maxwell scalar $\phi_1$ remains unchanged and also there is no effect on Vaidya mass function $f(u)$. Thus, if we consider $n$ radiations, considering the charge $e$ to be a function of $r$, the Einstein's field equations will imply that the mass of the embedded black hole will take the form
\begin{equation}
{\cal M}=M-(m_1 + m_2 + m_3 + m_4 + . . .+ m_n)+f(u)
\end{equation}
without affecting the Vaidya mass function $f(u)$ and the Maxwell scalar $\phi_{1}$. At some instant, there will
be an  thermal equilibrium state between $M$ and amount of radiation. If it still continues to radiate the equilibrium state will be disturbed. Taking Hawking's
radiation of charged black hole embedded in the rotating Vaidya
null radiating space, one might expect that the mass $M$ may be
radiated away, just leaving $M-(m_1 + m_2 + m_3 + m_4 + . . .+
m_n)$ equivalent to the Planck mass of about $10^{-5}\,{\rm g}$
and $f(u)$ untouched; that is, $M$ may not be exactly equal to
$(m_1 + m_2 + m_3 + m_4 +  . . . + m_n)$, but has a difference of
about a Planck-size mass. Otherwise, the mass $M$ may be
evaporated completely after continuous radiation, when $M=(m_1 +
m_2 + m_3 +m_4+ . . .+ m_n)$, just leaving the Vaidya mass
function $f(u)$ and the electric charge $e$ only.  Thus, we can
show this situation of the black hole in the form of a classical
space-time metric  without the mass $M$ as
\begin{eqnarray}
d s^2&=&[1-r^{-2}\{2rf(u)-e^2\}]\,du^2 +2du\,dr-r^2 d\theta^2 -r^2\rm sin^2\theta\,d\phi^2.
\end{eqnarray}
The Weyl scalar of this metric becomes
\begin{eqnarray}
\psi_2={1\over\,r^4}\Big\{e^2-r\,f(u)\Big\}.
\end{eqnarray}
and $\psi_3$, $\psi_4$ are unaffected. We observe
that the remaining metric (4.12) is the non-rotating charged Vaidya
null radiating black hole with $f(u)>e^2$ for $u=u_0$. The surface gravity
of the horizons at
\begin{equation}
r=r_\pm= f(u)\pm\surd{\{f(u)^2-e^2\}}
\end{equation}
is ${\cal K}_{\pm}=r^{-2}_{\pm}\{r_{\mp}-r_{\pm}\}$.
Consequently, the Hawking's temperature and entropy associated with the horizons are
$T_{\pm}$ = $(1/4\pi)\,r^{-2}_{\pm}\,\{r_{\mp}-r_{\pm}\}$ and  ${\cal S}=2\pi\,f(u)\{f(u)+\sqrt{f(u)^2
-e^2}\}-\pi e^2.$
If it goes on radiating, the mass of Reissner-Nordstrom becomes zero. But the presence of the Vaidya mass function $f(u)$ can avoid the formation of an ``instantaneous'' naked singularity with zero mass -- {\it a naked singularity that exists for an instant and then it continues its electrical radiation to create negative mass}. If we set the mass function $f(u)=0$ in (4.12), the metric will represent an `instantaneous' naked singularity with zero mass [1], and at that stage the gravity will depend only on electrical charge,
{\sl i.e.} $\psi_2={e^2/r^4}$, and not on the mass of black hole. However, the Maxwell scalar is unaffected which leads to the proof of non-rotating part of Theorem 5.

The time taken between two consecutive radiation is so short that one cannot physically realize how quickly radiation takes place. Immediately, after the exhaustion of the Reissner-Nordstrom mass, if one still continues to radiate electrically with $e(r)$, there may be formation of new mass $m^*_1$ (say). If this electrical radiation process continues further, the new mass will increase gradually as
\begin{equation}
{\cal M}^*=m^*_1+ m^*_2+ m^*_3+...  \,\;.
\end{equation}
This new mass will never decrease. Since the mass increases, the temperature will also start increasing from zero. Then, the  space-time metric will take the form
 \begin{eqnarray}
d s^2&=&[1+r^{-2}\{2r({\cal M}^*-f(u))+e^2\}]\,du^2 \cr
&&+2du\,dr-r^2d\theta^2 -r^2{\rm sin}^2\theta\,d\phi^2.
\end{eqnarray}

This metric will describe a black hole if $f(u)-{\cal M}^*>e^2$, that is,
when $f(u)>{\cal M}^*>e^2$. Thus, we have shown the changes in
the total mass of Reissner-Nordstrom-Vaidya black hole in classical
space-time metrics without affecting the Maxwell scalar $\phi_1$ and the
Vaidya mass function $f(u)$, for every electrical radiation during the
Hawking evaporation process. We have also observed
that, when $f(u)>{\cal M}^*$, the presence of Vaidya mass $f(u)$
in (4.16) can prevent the direct formation of negative mass naked
singularity. Otherwise, when $f(u)<{\cal M}^*$, this metric may
describe a `non-stationary' negative mass naked singularity, which
is different from the `stationary' one discussed in [1]. The
metric (4.16) can be written in Kerr-Schild ansatze in different backgrounds as
\begin{eqnarray*}
 g_{ab}^{\rm NMV}=g_{ab}^{\rm V}+2Q(r)\ell_a\ell_b,
\end{eqnarray*}
where $Q(r) =(r{\cal M}^*+e^2/2)r^{-2}$ in Vaidya background, and
\begin{eqnarray*}
g_{ab}^{\rm NMV}=g_{ab}^{\rm NM}+2Q(u,r)\ell_a\ell_b,
\end{eqnarray*}
with $Q(u,r) =-f(u)r^{-1}$. These Kerr-Schild forms show
that the  metric (4.1) is a solution of Einstein's field
equations and completes the proof of non-rotating part of Theorem 2 as mention in the introduction. Here the metric tensor $g_{ab}^{\rm V}$ is non-rotating
Vaidya null radiating metric, and $g_{ab}^{\rm NM}$ is that of the negative mass naked singularity with mass $\cal M^*$. The metric (4.16) leads to the  proof of the first part of Theorem 7.

\section{Changing mass of variable charged \\Kerr-Newman-Vaidya black hole}
\noindent
\setcounter{equation}{0}
\renewcommand{\theequation}{5.\arabic{equation}}
In order to show the Hawking's radiation in
the embedded Kerr-Newman-Vaidya black hole, we must have a different stress-energy
tensor as the Kerr-Newman-Vaidya black hole with $T_{ab}$
(3.2) does not have any direct Hawking's radiation effects.
The consideration of different forms of stress-energy-momentum
tensor in the study of Hawking's radiation effect in classical
space-time metrics is followed from Boulware's suggestion
[3] mentioned in introduction above.

It is noted that the Kerr-Newman-Vaidya black hole, describing the
Kerr-Newman black hole embedded into the rotating null
radiating Vaidya, is quite different from the standard
Kerr-Newman black hole. That is to say that, (i) the Kerr-Newman-Vaidya black
hole is algebraically special in Petrov classification with the
Weyl scalars $\psi_{0}=\psi_{1}=0$, $\psi_{2}= \psi_{3}=
\psi_{4}\neq 0$, where as the standard Kerr-Newman black hole is
Petrov type $D$ with $\psi_{0}=\psi_{1}=\psi_{3}=\psi_{4}=0$,
$\psi_{2} \neq 0$. (ii) the Kerr-Newman-Vaidya black hole
possesses the total energy-momentum tensor (3.2), representing
interaction of the null fluid $T^{(\rm n)}_{ab}$ with
the electromagnetic field $T^{(\rm E)}_{ab}$, {\it i.e.}, the charged
null radiating fluid; however the energy-momentum tensor of the
Kerr-Newman black hole is that of electromagnetic field $T^{(\rm
E)}_{ab}$, simply a charged black hole. (iii) the
Kerr-Newman-Vaidya black hole is a non-stationary extension of the
stationary Kerr-Newman black hole. Due to the above differences
between the two black holes, it is worth to study the physical
properties of the embedded black hole. Hence, it is hoped that the
study of Hawking electrical radiation in the embedded
Kerr-Newman-Vaidya black hole will certainly lead to a different
physical feature than that of {\sl non-embedded} Kerr-Newman black
hole.

Thus, we consider the line element of a variable-charged
Kerr-Newman-Vaidya black hole with the charge $e$ varying with respect to the coordinate $r$ in (3.1) as follows:
\begin{eqnarray}
d s^2&=&[1-R^{-2}\{2r(M+f(u))- e^2(r)\}]\,du^2 +2du\,
dr \cr &&+2aR^{-2}\{2r(M+f(u))-e^2(r)\}\,{\rm sin}^2
\theta\,du\,d\phi -2a\,{\rm sin}^2\theta\,dr\,d\phi \cr
&&-R^2d\theta^2 -\{(r^2+a^2)^2-\Delta^*a^2\,{\rm
sin}^2\theta\}\,R^{-2}{\rm sin}^2\theta\,d\phi^2,
\end{eqnarray}
where $\Delta^*=r^2-2r\{M+f(u)\}+a^2+e^2(r)$. Then, with the
variable charge $e(r)$, there will be a
different energy-momentum tensor from that of the original
Kerr-Newman-Vaidya metric (3.1). Accordingly, we have calculated the Ricci and Weyl scalars from the field equations with
the functions $e(r)$ and found as follows:
\begin{eqnarray}
&\phi_{11}=&{1\over
{2\,R^2\,R^2}}\,\Big\{e^2(r)-2r\,e(r)e(r)_{,r}\Big\}
\cr  &&+{1\over 4R^2}\,\Big\{e^2(r)_{,r}+e(r)\,e(r)_{,rr}\Big\}, \\
&\phi_{22}=&-{r\over 2\,R^2\,R^2}\Big\{2\,r\,f(u)_{,u}+a^2{\rm
sin}^2\theta\,f(u)_{,uu}\Big\}, \cr &\phi_{12}=&{i\,a\,{\rm
sin}\,\theta\,\over{\surd 2\,K\,\bar{R}\,R^2}}\,f(u)_{,u},
\nonumber \\ &\Lambda =& -{1\over
12\,R^2}\,\Big\{e^2(r)_{,r}+e(r)\,e(r)_{,rr}\Big\}, \\
&\psi_2=&{1\over\bar{R}\,\bar{R}\,R^2}\Big[e^2(r)-R\{M+f(u)\} \cr &&+{1\over 6}\,\bar{R}\,\bar{R}\,\Big\{e^2(r)_{,r}
+e(r)\,e(r)_{,rr}\Big\}-\bar{ R}\,e(r)\,e(r)_{,r}\Big], \\
&\psi_3=&{-i\,a\,{\rm sin}\theta\over 2\surd 2\bar{R}\,\bar{ R}\,R^2}\Big\{(4\,r+\bar{R})f(u)_{,u}\Big\}, \cr
&\psi_4=&{{a^2r\,{\rm sin}^2\theta}\over 2 \bar{R}\,\bar{
R}\,R^2\,R^2}\,\Big\{R^2f(u)_{,uu}-2rf(u)_{,u}\Big\}. \nonumber
\end{eqnarray}
From these we observe that the variable charge $e(r)$ has the
effect only on the scalars $\phi_{11}$, $\Lambda$ and $\psi_2$.
However, the non-vanishing of $\phi_{22}$, $ \phi_{12}$,
$\psi_3$ and $\psi_4$ indicate the different physical feature of
non-stationary Kerr-Newman-Vaidya black hole from the stationary
standard Kerr-Newman. One important feature of the
field equations corresponding to the metric (5.1) with $e(r)$ is
that the expressions for $\phi_{11}$ and $\Lambda$ do not involve
the Vaidya mass function $f(u)$. So it suggests the possibility to
study the Hawking's electrical radiation in this non-stationary embedded
black hole.

Now for an electromagnetic field, the Ricci scalar $\Lambda\equiv
(1/24)g^{ab}R_{ab}$ in (5.3) has to vanish leading to the solution
\begin{equation}
e^2(r)=2rm_1 + C
\end{equation}
where $m_1$ and $C$ are real constants. Substituting this
$e^2(r)$ in (5.2) and (5.4), and after identifying the
constant $C\equiv e^2$, we obtain
\begin{eqnarray}
&&\phi_{11}={e^2\over {2\,R^2\,R^2}}, \\
&&\psi_2={1\over{\bar{R}\,\bar{R}\,R^2}}\Big[e^2-R\{M-m_1+f(u)\}\Big],
\end{eqnarray}
and $\phi_{12}$, $\phi_{22}$, $\psi_3$ and $\psi_4$ are remained
unaffected as in (5.2) and (5.4) above. Accordingly, the Maxwell
scalar $\phi_1={1\over 2}F_{ab}(\ell^a\,n^b+\overline{m}^a\,m^b)$ with the constant charge $e$ becomes
\begin{eqnarray}
\phi_{1}={e\over {\surd 2\,\bar{R}\,\bar{R}}},
\end{eqnarray}
which is the same Maxwell scalar of the Kerr-Newman-Vaidya
metric (3.1) if once written out from (B.3) below. From the Weyl scalar
(5.7) we have clearly seen {\sl a change in the mass $M$} by some
constant quantity $m_1$  for the first  electrical
radiation in the embedded black hole. Then the total mass of
Kerr-Newman-Vaidya black hole becomes $(M-m_1)+f(u)$ and the
metric takes the form
\begin{eqnarray}
ds^2&=&[1-R^{-2}\{2r(M-m_1 +f(u))-e^2\}]\,du^2 +2du\,dr \cr &&+2aR^{-2}\{2r(M-m_1+f(u))-e^2\}\,{\rm
sin}^2\theta\,du\,d\phi-2a\,{\rm sin}^2\theta\,dr\,d\phi\cr &&-R^2d\theta^2 -\{(r^2+a^2)^2 -\Delta^*a^2\,{\rm
sin}^2\theta\}\,R^{-2}{\rm
sin}^2\theta\,d\phi^2,
\end{eqnarray}
where $\Delta^*=r^2-2r\{M-m_1+f(u)\}+a^2+e^2$. Since the Maxwell
scalar (5.8) remains the same for the first Hawking radiation, we
again consider the charge $e$ to be a function of $r$ for the
second radiation in the metric (5.9) with the mass $M-m_1+f(u)$.
This process will certainly lead, by the Einstein-Maxwell field equations
with the vanishing of $\Lambda$, to reduce another quantity $m_2$
(say) from the total mass, {\it i.e.} the mass becomes
$M-(m_1+m_2)+f(u)$ after the second electrical radiation. Here
again, we observe that the Maxwell scalar $\phi_1$ remains the
same form and also there is no effect on the Vaidya mass
$f(u)$ after the second radiation. Thus, if we consider $n$
radiations, every time taking the charge $e$ to be a function of
$r$, the Einstein's field equations will imply that the total mass
of the black hole takes the form
\begin{equation}
{\cal M}=M-(m_1 + m_2 + m_3 + m_4 + . . .+ m_n)+f(u)
\end{equation}
without affecting the mass function $f(u)$. Taking Hawking's
radiation of charged black hole embedded in the rotating Vaidya
null radiating space, one might expect that the mass $M$ may be
radiated away, just leaving the expression $M-(m_1 + m_2 + m_3 + m_4 + . . .+
m_n)$ equivalent to the Planck mass of about $10^{-5}\,{\rm g}$
and $f(u)$ untouched; that is, $M$ may not be exactly equal to
$(m_1 + m_2 + m_3 + m_4 +  . . . + m_n)$, but has a difference of
about a Planck-size mass. Otherwise, the mass $M$ may be
evaporated completely after continuous radiation, when $M=(m_1 +
m_2 + m_3 +m_4+ . . .+ m_n)$, just leaving the Vaidya mass
$f(u)$ and the electric charge $e$ only.  Thus, this situation of the black hole without Kerr mass $M$ may be seen in the form of a classical
space-time metric as
\begin{eqnarray}
d s^2&=&[1-R^{-2}\{2r f(u)-e^2\}]\,du^2 +2du\,dr \cr
&&+2aR^{-2}\{2rf(u)-e^2\}\,{\rm sin}^2 \theta\,du\,d\phi-2a\,{\rm sin}^2\theta\,dr\,d\phi \cr &&-R^2d\theta^2 -\{(r^2+a^2)^2
-\Delta^*a^2\,{\rm sin}^2\theta\}\,R^{-2}{\rm
sin}^2\theta\,d\phi^2,
\end{eqnarray}
where $\Delta^*=r^2-2rf(u)+a^2+e^2$. The Weyl scalar of this
metric becomes
\begin{eqnarray}
\psi_2={1\over\bar R\,\bar
R\,R^2}\Big\{e^2-R\,f(u)\Big\},
\end{eqnarray}
and $\psi_3$, $\psi_4$ are unaffected as (5.4). From (5.11) we know
that the remaining metric is the rotating charged Vaidya
null radiating black hole with $f(u)>a^2+e^2$. The surface gravity for the solution
at the horizon
\begin{equation}
r=r_{\pm}= f(u)\pm\surd{\{f(u)^2-a^2-e^2\}},
\end{equation}
is
${\cal K}_{\pm}=(1/2)R^{-2}_{\pm}\{r_{\mp}-r_{\pm}\}$.
The Hawking's temperature on the horizons are
$T_{\pm}=(1/4\pi) R^{-2}_{\pm}\{r_{\mp}-r_{\pm}\}$.
The entropy and angular velocity of the horizon are
found as
\begin{eqnarray*}
&&{\cal S}=2\pi\,f(u)\Big\{f(u)+\sqrt{f(u)^2
-(a^2+e^2)}\,\Big\}-\pi e^2, \cr
&&\Omega_{\pm}=\frac{a\{2rf(u)-e^2\}}{(r^2+a^2)^2}\Big|_{r=r_{+}}.
\end{eqnarray*}
Thus, we may regard this left out remnant of the Hawking
evaporation as the rotating charged Vaidya black hole. On the
other hand, the metric (5.11)
may be interpreted that the presence of Vaidya mass function $f(u)$
can avoid the formation of an `instantaneous' naked singularity
with zero mass. The formation of `instantaneous' naked singularity
with zero mass in {\sl non-embedded} Reissner-Nordstrom and
Kerr-Newman, black holes is unavoidable during Hawking's
evaporation process, as shown in [1]. That is, if we set the mass
function $f(u)=0$, the metric (5.11) would certainly represent an
`instantaneous' naked singularity with zero mass, and at that
stage gravity of the surface would depend only on electric charge,
{\it i.e.} $\psi_2=(e^2/{\bar{R}\,\bar{R}\,R^2)}$, and not
on the mass of black hole. However, the Maxwell scalar $\phi_1$ is
unaffected. Thus, from (5.11) with $f(u)\neq 0$ it seems natural
to refer to the {\it rotating} charged Vaidya null radiating black
hole as an `instantaneous' black hole, during the Hawking's evaporation process of Kerr-Newman-Vaidya black hole which proves the rotating part of Theorem 4.

   The time taken between two consecutive radiations is supposed
to be so short that one may not physically realize how quickly
radiations take place. Immediately  after the exhaustion of the
Kerr-Newman mass, if one continues the remaining solution
(5.11) to radiate electrically with $e(r)$, there may be a
formation of a new mass $m^*_1$ (say). If this electrical radiation
process continues forever, the new mass will increase gradually as
\begin{equation}
{\cal M}^*=m^*_1 + m^*_2 + m^*_3 + m^*_4 + . . .   \,\; .
\end{equation}
However, it appears that this new mass will never decrease.
Then, the space-time metric will take the following form
\begin{eqnarray}
d s^2&=&[1+R^{-2}\{2r({\cal M}^*-f(u))+e^2\}]\,du^2 +2du\,dr \cr
&&+2aR^{-2}\{2r(f(u)-{\cal M}^*)-e^2\}\,{\rm sin}^2\theta\,du\,d\phi-2a\,{\rm sin}^2\theta\,dr\,d\phi\cr &&
-R^2d\theta^2 -\{(r^2+a^2)^2 -\Delta^*a^2\,{\rm
sin}^2\theta\}\,R^{-2}{\rm sin}^2\theta\,d\phi^2,
\end{eqnarray}
where $\Delta^*=r^2-2r\{f(u)-{\cal M}^*\}+a^2+e^2$. This metric
will describe a black hole if $f(u)-{\cal M}^*>a^2+e^2$, that is,
when $f(u)>{\cal M}^*>a^2+e^2$ for a particular value of $u$. Thus, we have shown the changes in
the total mass of Kerr-Newman-Vaidya black hole in classical
space-time metrics without affecting the Maxwell scalar and the
Vaidya mass, for every electrical radiation during the
Hawking evaporation process. We have also observed
that, when $f(u)>{\cal M}^*$, the presence of Vaidya mass $f(u)$
in (5.15) can prevent the direct formation of negative mass naked
singularity. Otherwise, when $f(u)<{\cal M}^*$, this metric may
describe a `non-stationary' negative mass naked singularity, which
is different from the `stationary' one discussed in [1]. The
metric (5.15) can be expressed in Kerr-Schild ansatze in different backgrounds as
\begin{eqnarray}
 g_{ab}^{\rm NMV}=g_{ab}^{\rm V}+2Q(r,\theta)\ell_a\ell_b,
\end{eqnarray}
where $Q(r,\theta) =(r{\cal M}^*+e^2/2)R^{-2}$, and
\begin{eqnarray}
g_{ab}^{\rm NMV}=g_{ab}^{\rm NM}+2Q(u,r,\theta)\ell_a\ell_b,
\end{eqnarray}
with $Q(u,r,\theta) =-rf(u)R^{-2}$. These Kerr-Schild forms show
that the  metric (5.15) is a solution of Einstein's field
equations. Here the metric tensor $g_{ab}^{\rm V}$ is rotating
Vaidya null radiating black hole metric, and $g_{ab}^{\rm NM}$ is the metric
describing the rotating negative mass naked singularity. The metric (5.15) leads to the proof of Theorem 7 for creating negative mass naked singularity of the rotating part, as mentioned in the introduction.

\section{Conclusion}
In this paper, by adopting Wang-Wu mass function we derive a class of embedded exact solutions of Einstein's field equations, namely non-rotating Reissner-Nordstrom-Vaidya  and rotating Kerr-Newman-Vaidya solutions. The gravitational structure of the solutions are analyzed by observing the nature of the energy momentum tensor. The energy-momentum tensors (2.4) and (3.2) associated with the embedded solutions obey the energy conservation equations $T^{ab}_{\;\;\;\,;b}=0$, showing that the space-time metrics (2.3) and (3.1) are exact solutions of field equations. These tensors show the interaction of Vaidya null fluid $T^{(\rm n)}_{ab}$ with the electromagnetic field $T^{(\rm E)}_{ab}$ -- no-rotating in (2.4) and rotating in (3.2).

These embedded non-rotating Reissner-Nordstrom-Vaidya and rotating Kerr-Newman-Vaidya solutions can also be able to expressed in Kerr-Schild ansatze. The solution of  Reissner-Nordstrom-Vaidya is  of Petrov type D; and Kerr-Newman-Vaidya solution is  algebraically special in Petrov classification of space-time metric. In embedded   Reissner-Nordstrom-Vaidya  and Kerr-Newman-Vaidya black hole, the presence of Vaidya mass completely prevent the disappearance of black hole masses during the radiation process and thereby, the formation of ``instantaneous'' charged Vaidya black holes, non-rotating in (4.12) and rotating in (5.11). However, the non-embedded Reissner-Nordstrom and Kerr-Newman cannot prevent the disappearance of the total mass of the evaporating black holes.

It appears that (i) the changes in the mass of black holes, (ii)
the formation of `instantaneous' naked singularities  with zero
mass and (iii) the creation of `negative mass naked singularities'
in {\sl non-embedded} Reissner-Nordstrom as well as Kerr-Newman
black holes [1] are presumably the correct formulation in
classical space-time metrics of the three possibilities of black
hole evaporation suggested by Hawking and Israel [16]. However,
the creation of `negative mass naked singularities' may be a
violation of Pentose's cosmic censorship hypothesis [19]. The embedded black
holes discussed here can be expressed in Kerr-Schild ansatze, accordingly their consequent
negative mass naked singularities are also expressible in
Kerr-Schild forms showing them as solution of Einstein's field
equations. It is also observed that once a charged black hole is
embedded into some spaces, it will continue to embed into the same space forever
through out its Hawking evaporation process. For example,
Kerr-Newman black hole is embedded into the rotating Vaidya null
radiating universe, it continues to embed as `instantaneous'
charged Vaidya black hole in (5.11) and embedded negative mass naked
singularity as in (5.15). There Hawking's radiation does not
affect the Vaidya mass through out the evaporation process of
Kerr-Newman mass.  This means that the embedded negative mass
naked singularities (5.15) possess the total
energy momentum tensors (3.2), as
the Kerr-Newman mass does not involved in these tensors (3.3) and (3.4), and the
change in the mass due to continuous radiation does not affect
them. Thus, it may be concluded that once a black hole is embedded
into some spaces, it will continue to embed forever without
disturbing the nature of matters present in the backgrounds. Also since there is no effect on the Vaidya mass $f(u)$ during Hawking's
evaporation process, it will always remain unaffected. That is to say that,
unless some external forces apply to remove the mass $f(u)$
from the embedded space-time geometries, it will
certainly continue to exist along with the electrically radiating
black holes, rotating or non-rotating forever. If one accepts the
Hawking continuous evaporation of charged black holes, the loss of
the mass and creation of a new mass are the process of the continuous
radiation. So, from the space-time metrics (5.9), (5.11), (5.15),
it may also be concluded that once electrical
radiation starts, it will continue to radiate forever describing
the various stages of the life of radiating black holes. This establishes the proof of the rotating part of the Theorem 8 and simultaneously follows the non-rotating case of the theorem.

Also, we find from the above that the change in the mass of embedded black
holes, takes place due to the Maxwell scalar ${\phi_1}$, remaining unchanged in the field equations during continuous radiation. So, if the Maxwell scalar
${\phi_1}$ is absent from the space-time geometry, there will be
no such electrical radiation, and consequently, there will be no observable changes in the mass of the black hole solution. Therefore, we cannot, theoretically,
expect to observe  such {\sl relativistic change} in the mass of
uncharged Schwarzschild as well as Kerr black holes. This suggests
that these uncharged Schwarzschild as well as Kerr black holes
will forever remain the same without changing their life styles.

From the study of Hawking's radiation above, it is also
found that, as far as the embedded black holes are concerned,
the Kerr-Newman black hole has relations with other
rotating black holes, like the charged Vaidya black hole here. There the later ones are `instantaneous' black holes of the
respective embedded ones. It is observed that the classical space-time
metrics discussed above would describe the possible life style of
radiating embedded black holes at different stages during their
continuous radiation. These {\sl embedded} classical space-time metrics
describing the changing life style of black holes are different from the
{\sl non-embedded} ones studied in [1] in various respects shown
above. Here the study of these embedded solutions suggests the
possibility that in an early universe there might be some black
holes, which might have embedded into some other spaces possessing
different matter fields with well-defined physical properties.

\section*{Acknowledgement}
Ibohal acknowledges his appreciation for hospitality received from Inter-University Centre for Astronomy and Astrophysics (IUCAA), Pune during his visit in preparing the paper. This work is supported by UGC Major Research Project, reference No.F. 31-87/2005(SR) dated 31st March 2006.

\section*{Appendix A: Energy-momentum tensor and energy conservation equations}
\setcounter{equation}{0}
\renewcommand{\theequation}{A.\arabic{equation}}

Here we present a rotating metric having a mass function $\hat{M}(u,r)$ given in Ref. [2] equation (6.4), which has been utilized in the derivation of the embedded Reissner-Nordstrom-Vaidya and Kerr-Newman-Vaidya solutions.
\begin{eqnarray}
d s^2&=& \{1-2r\hat{M}(u,r)R^{-2}\}\,du^2+2du\,dr \cr && +4a\,r\,\hat{M}(u,r)R^{-2}\sin^2\theta\,du\,d\phi -2a\,\sin^2\theta\,dr\,d\phi \cr &&-R^2d\theta^2 -\{(r^2+a^2)^2 -\Delta a^2\,{\rm sin}^2\theta\}R^{-2}\sin^2\theta\,d\phi^2,
\end{eqnarray}
where $\Delta = r^2-2r\hat{M}(u,r)+a^2$ and $R^2 = r^2 + a^2\cos^2\theta$. When $a=0$, this metric will reduce to a non-rotating canonical metric in Eddington-Finkestein coordinate system,
\begin{eqnarray}
ds^2=\Big\{1-\frac{2}{r}\hat{M}(u,r)\Big\} du^2+2du\,dr-r^2(d\theta^2+{\rm sin}^2\theta\,d\phi^2).
\end{eqnarray}
By virtue of Einstein's field equations associated with the above line element (A.1) we obtain the total energy momentum tensor (EMT) as
follows:
\begin{eqnarray}
T_{ab} &=& T^{(\rm n)}_{ab} +T^{(\rm m)}_{ab} \cr
&=&\mu^*\,\ell_a\,\ell_b+
2\,\rho^*\,\ell_{(a}\,n_{b)}
+2\,p\,m_{(a}\bar{m}_{b)}
 + 2\,\omega\,\ell_{(a}\,\bar{m}_{b)} +
2\,\bar\omega\,\ell_{(a}\,m_{b)}
\end{eqnarray}
where the EMTs for the rotating null fluid as well as that of the
rotating matter are respectively given below:
\begin{eqnarray}
&&T^{(\rm n)}_{ab}= \mu^*\,\ell_a\,\ell_b +
2\omega\,\ell_{(a}\,\bar{m}_{b)}+2\bar\omega\,\ell_{(a}\,m_{b)}\cr
&&T^{(\rm m)}_{ab}=2\,(\rho^*+p)\,\ell_{(a}\,n_{b)}
- p\,g_{ab} ,
\end{eqnarray}
where $\bar\omega$ is the complex conjugate of $\omega$ referred to as a rotational density. When
$a=0$, the rotational density $\omega$ will be zero, and then these EMTs are similar to those introduced by Husain [27] in the case of non-rotating fluid. Here the quantities $\mu^*$, $\rho^*$. $p$ and $\omega^*$ in (A.3) are found as
\begin{eqnarray}
&&K\mu^*=-{1\over R^2\,R^2}\,\Big\{2r^2\hat{M}(u,r)_{,u} + a^2r\sin^2\theta \hat{M}(u,r)_{,uu}\Big\} \cr
&&K\rho^* =  \frac{2r^2}{R^2\,R^2}\hat{M}(u,r)_{,r}\cr
&&Kp=  {1\over R^2\,R^2}\Big\{2\,r^2\hat{M}(u,r)_{,r}-R^2\Big(2\hat{M}(u,r)_{,r}
+r\,\hat{M}(u,r)_{,rr}\Big)\Big\} \cr
&&K\omega =-\frac{i\,a\,\sin\theta}{\surd 2\,R^2\,R^2}\, \Big\{R\,\hat{M}(u,r)_{,u}-
r\,\bar R\,\hat{M}(u,r)_{,ru}\Big\}
\end{eqnarray}
with the universal constant $K=8\pi G/c^4$. The trace of $T^{(\rm n)}_{ab}$ is zero and that of $T^{(\rm m)}_{ab}$ is $T=2(\rho^*-p)$. The line element (A.1) contains many known spherically rotating axisymmetric solutions.

The energy conservation equations for the energy-momentum tensor (A.2) $T^{ab}_{\;\;\;\,;b}=0$ can, in general, be transcribed in Newman-Penrose spin coefficient formalism as follows
\begin{eqnarray}
D\rho^{*}&=&\rho^{*}(\rho+\bar{\rho})+p(\rho+\bar{\rho})+\omega
\bar{\kappa}+ \bar{\omega} \kappa  \\
D\mu^*+\nabla \rho^* + \bar{\delta}\omega + \delta \bar{\omega}&=& \mu^*\{(\rho+\bar{\rho})- 2 (\epsilon+\bar{\epsilon})\} -\rho^{*}(\mu+\bar{\mu}) - p (\mu+\bar{\mu}) \cr &&- \omega(2\pi+ 2\bar{\beta}- \bar{\tau}) -\bar{\omega}(2\bar{\pi}+ 2\beta- \tau)\\
D\bar{\omega}+ \bar{\delta}p&=&\mu^*\bar{\kappa}+\rho^*(\bar{\tau}-\pi)+ p(\bar{\tau}-\pi)
+\omega \bar{\sigma}\cr &&- \bar{\omega}(2\epsilon-2\bar{\rho}-\rho)
\end{eqnarray}
where $\kappa, \tau, \pi$ etc are spin coefficients and the derivative operators are defined as
\begin{equation}
D\equiv \ell^a\partial_a, \;\,\nabla \equiv n^a \partial_a,\;\,\delta \equiv m^a \partial_a, \;\,\bar{\delta} \equiv \bar{m}^a \partial_a.
\end{equation}
These (A.6-A.8) are general equations for the energy conservation laws associated with any energy-momentum tensor of the type (A.3). An energy-momentum tensor (A.3) for any rotating solution has to satisfy these energy conservation equations; then the metric can to be called an exact solution of Einstein's equations.

\vspace*{0.15in}
\section*{Appendix B: NP spin coefficients for the Kerr-Newman-Vaidya black hole}
\setcounter{equation}{0}
\renewcommand{\theequation}{B.\arabic{equation}}
\vspace*{0.15in}
For the analysis of the embedded Kerr-Newman-Vaidya black hole (3.1), we cite the complex null tetrad vectors for the line element as follows
\begin{eqnarray}
&&\ell_a=\delta^1_a -a\,{\rm sin}^2\theta\,\delta^4_a,\cr
&&n_a=\frac{\Delta}{2\,R^2}\,\delta^1_a+ \delta^2_a
-\frac{\Delta}{2\,R^2}\,\,a\,{\rm sin}^2\theta\,\delta^4_a,\\
&&m_a=-\frac{1}{\surd 2R}\,\Big\{-ia\,{\rm
sin}\,\theta\,\delta^1_a+R^2\,\delta^3_a +i(r^2+a^2)\,{\rm
sin}\,\theta\,\delta^4_a\Big\},\nonumber
\end{eqnarray}
where $\Delta=r^2-2r\{M+f(u)\}+a^2+e^2$.
Using these null tetrad vectors, we solve the Cartan's first structure equations, transcribed in NP formalism by McIntosh and Hickman [19], for the spin coefficients which are found as follows
 \begin{eqnarray}
&&\kappa=\epsilon=\lambda=\sigma=0 , \cr
&&\rho=-\frac{1}{\bar R},\quad
\mu=-\frac{\Delta}{2 {\bar R}R^2},\cr
&&\alpha={(2ai-R\,{\rm cos}\,\theta)\over{2\surd 2\bar
R\,\bar R\,{\rm sin}\,\theta}},\quad \beta={{\rm
cot}\,\theta\over2\surd 2R},\\
&&\pi={i\,a\,{\rm sin}\,\theta\over{\surd 2\bar R\,\bar
R}},\quad \tau=-{i\,a\,{\rm sin}\,\theta\over{\surd 2R^2}},\cr
&&\gamma={1\over{2\bar R\,R^2}}\,\left[\{r-m-f(u)\}\bar
R-\Delta\right].\nonumber
\end{eqnarray}
The tetrad components of Ricci and Weyl tensors, known as Ricci and Weyl scalars respectively are obtained from the Cartan's second structure equations.
The Ricci scalars are as follows
\begin{eqnarray}
&&\phi_{00}=\phi_{01}=\phi_{10}=\phi_{20}=\phi_{02}=\Lambda=0 \cr
&&\phi_{11}=\frac{e^2}{2 \,R^2\,R^2}\cr
&&\phi_{12}=\frac{i a\, {\rm sin\,\theta}}{2\surd 2 R^2 \bar R}f(u)_{,u}\\
&&\phi_{22}=-\frac{r}{2K \,R^2\, R}\{2rf(u)_{,u}+a^2\sin^2\theta f(u)_{,uu}\}.\nonumber
\end{eqnarray}
Here it is found that the Ricci scalar $\Lambda=(1/24)g^{ab}R_{ab}$ vanishes for the rotating solution (3.1). This shows the characteristic feature of Kerr-Newman-Vaidya that it is an electrically charged black hole. The Maxwell scalar $\phi_1$ can be obtained from the Ricci scalar $\phi_{11}$ using the relation $\phi_{11}=\phi_1 \bar{\phi}_1$ for electromagnetic field $F_{ab}$ [17].
The Weyl scalars  become
\begin{eqnarray}
&&\psi_0=\psi_1=0\crcr
&&\psi_2=-\frac{1}{\bar R\,\bar R\,R^2}\Big[R\,\{m+f(u)\}-e^2\Big] \cr
&&\psi_3 =-{i\,a\,{\rm sin}\theta\over
2\surd 2\bar R\,\bar R\,R^2}\Big\{(4\,r+\bar
R)f(u)_{,u}\Big\}  \\
&&\psi_4 ={{a^2r\,{\rm sin}^2\theta}\over 2\bar R\,\bar
R\,R^2\,R^2}\,\Big\{R^2f(u)_{,uu}-2rf(u)_{,u}\Big\}. \nonumber
\end{eqnarray}
The vanishing of Weyl scalars $\psi_0$ and $\psi_1$ indicates that the Weyl curvature tensor of  the solution is algebraically special type II in Petrov classification with a repeated principal null congruence $\ell_a$ (2.4), which is geodesic $(\kappa=\epsilon=0)$, shear free $(\sigma=0)$, expanding [$2\,\hat{\theta}\equiv
\ell^a_{\,;a}= -(\rho + \bar{\rho})$] as well as non-zero twist
[$4\,\hat{\omega}^2\equiv 2\ell_{[a;\,b]}\ell^{a;\,b}=-(\rho - \bar{\rho})$]. Here $\kappa, \epsilon, \sigma, \rho$ are NP spin coefficients given in (B.2). This completes the proof of the Theorem 2 stated in the introduction above. For future use we also introduce
the Ricci and Riemann Curvature invariants for the metric (3.1) as
\begin{eqnarray}
R_{ab}R^{ab}&=&\frac{4e^4}{R^2R^2R^2R^2}.\cr
R_{abcd}R^{abcd}&=&24\Big[\frac{(R\{m+f(u)\}-e^2)^2}{R^2R^2\bar R \bar R\bar R\bar R}\cr &&+\frac{({\bar R}\{m+f(u)\}-e^2)^2}{R^2R^2 R R R R}\Big]+\frac{20 e^4}{R^2 R^2 R^2 R^2}.
\end{eqnarray}

Here we shall verify that the components of $T^{ab}$ with the quantities $\mu^*, \rho^*, p$ and $\omega$ given in (3.5a-c) for the rotating Kerr-Newman-Vaidya metric satisfy the conservation equations (A.6-A.8).
The  directional derivative operators (A.9) along the null vectors (B.1) are given as follows:
\begin{eqnarray}
&&D=\partial_r, \cr
&&\nabla=\frac{1}{R^2}\{(r^2+a^2)\,\partial_u -\frac{\Delta}{2} \,\partial_r +a\, \partial_\phi\}, \cr
&&\delta=\frac{1}{\sqrt{2}R}\{i a\, sin\,\theta \,\partial_u +\partial_\theta +\frac{i}{sin \, \theta}\,\partial_\phi\},\cr
&&\bar{\delta}=\frac{1}{\sqrt{2}\bar{R}}\{-i a \,sin\,\theta \,\partial_u +\partial_\theta -\frac{i}{sin \,\theta}\,\partial_\phi\}.
\end{eqnarray}
It is to say that the equations (A.6) and (A.8) are comparatively easier to verify than (A.7). Now, by virtue of (3.5) and (B.2), the left side of (A.7) is found as
\begin{eqnarray}
\lefteqn{D\mu^*+\nabla \rho^*+\bar{\delta}\omega+\delta\bar{\omega}} \cr\cr
&=&\frac{2 r e^2\Delta}{K R^2 R^2 R^2 R^2}+\frac{2 r^2a^2\sin^2\theta}{K R^2 R^2 R^2}f(u)_{,uu} +\frac{2r}{K R^2 R^2 R^2}\Big\{2r^2-a^2\cos^2\theta\Big\}f(u)_{,u} \cr
&&+\frac{r f(u)_{,u}}{K R^2 R^2 R^2 R^2}\Big\{a^2 sin^2\theta (7 a^2 \cos^2\theta- r^2)\Big\},
\end{eqnarray}
which can be shown equal to the right side of (A.7), by using spin coefficients (B.2).
It indicates the fact that the energy-momentum tensor (3.2) satisfies the conservation equation (3.6). Similarly, one can verify the conservation laws for the case of the non-rotating Reissner-Nordstrom-Vaidya metric (2.3) having $T_{ab}$ (2.7), when $a=0$. This shows the fact that the Reissner-Nordstrom-Vaidya and Kerr-Newman-Vaidya are solutions of Einstein's field equations.


\begin{thebibliography}{99}
\bibitem{x}Ibohal, N. (2002), ``On the variably-charged black holes in general relativity: Hawking's radiation and naked singularities'', {\sl Class. Quantum Grav.} {\bf 19}, 4327-4341, gr-qc/0405019.
\bibitem{x}Hawking, S. W. (1974), ``Black Hole explosion'', {\sl Nature (Lond.)} {\bf 248}, 30-31; (1975), ``Particle creation by black holes'', {\sl Commun. math. Phys.} {\bf 43}, 199-220; (1976), ``Break down of predictability in gravitational collapse'', {\sl Phys. Rev. D} {\bf 14}, 2460-2470.
\bibitem{x}Boulware D. G. (1976), ``Hawking radiation and thin shells'', {\sl Phys. Rev. D}, {\bf 13}, 2169-2187.
\bibitem{x}Steinmular, B., King, A. R. and Losota, J. P. (1975), ``Radiating bodies and naked singularities'', {\sl Phys Lett} {\bf 51A}, 191-192.
\bibitem{x}Tipler,F. J., Clerke , C. J. S. and  Ellis, G. F. R. (1980)
`Singularities and Horizons -- A review article' in {\sl
General Relativity and Gravitation: One Hundred Year After the birth of
Albert Einstein}, Vol 2, edited by A. Held (Plenum Press, New
York).
\bibitem{x}Guth, A. H. (1981), ``Inflationary universe: A possible solution to the horizon and flatness problems'', {\sl Phys. Rev D}, {\bf 23}, 347-356.
\bibitem{x}Ibohal, N. (2005), ``On the variable-charged Black holes Embedded into de Sitter space: Hawking's Radiation'', {\sl Int. J. Mod. Phys. D} {\bf 14}, 973-994, gr-qc/0405018.
\bibitem{x}Carmeli, M. and Kaye, M. (1977), ``Gravitational field of a radiating rotating body'',  {\sl Ann. Phys. (N.Y.)},  {\bf 103}, 97-120.
\bibitem{x}Carmeli, M. (1983), {\it  Classical Fields, General Relativity and Gauge Theory} ( John Wiley,  New York).
\bibitem{x}Ibohal, N. (2005), ``Rotating metrics admitting non-perfect fluids'', {\sl Gen. Relativ. Gravit.} {\bf 37}, 19-51, gr-qc/0403098.
\bibitem{x}Vaidya, P. C. (1999), ``The Gravitational field of a Rotating Star'',  {\sl Proc. Indian Acad. Sci.}  {\bf A33}, 264 (1951); reprinted {\sl Gen. Relativ. Gravit.} {\bf 31}, 119-135.
\bibitem{x}Cai, R. G., Ji, J. Y. and Soh, K. S. (1998), ``Action and entropy of black holes in spacetimes with a cosmological constant'', {\sl Class Quantum Grav.} {\bf 15}, 2783-2793.
\bibitem{x}Glass, E. N. and Krisch, J. P. (1998), ``Radiation and String atmosphere for relativistic stars'',  {\sl Phys. Rev. D} {\bf 57}, R5945-5947; (1999), ``Two-fluid atmosphere for relativistic stars'', {\sl Class. Quantum Grav.} {\bf 16}, 1175-1184.
\bibitem{x}Xanthopoulos, B. C. (1978), ``Exact vacuum solutions of Einstein's equation from linearized solutions'', {\it J. Math. Phys.} {\bf 19}, 1607-1609.
\bibitem{x}Wang, A. and Wu, Y. (1999), ``Generalized Vaidya solutions'', {\sl Gen. Relativ. Gravit.} {\bf 31}, 107-114.
\bibitem{x}Hawking, S. H. and Israel, W. (1980), ``An Introductory survey'' in {\sl General Relativity: An Einstein Centenary Survey} eds. by S.W. Hawking  and W. Israel  (Cambridge Univ. Press, Cambridge), pp 1-23.
\bibitem{x}Newman, E. T. and Penrose, R. (1962), ``An approach to gravitational radiation by a method of spin coefficients'', {\sl J. Math. Phys.} {\bf 3}, 566-578.
\bibitem{x} Bekenstein, J.D. (1973), ``Black holes and Entropy'', {Phys. Rev. D.} {\bf 7}, 2333-3300.
\bibitem{x}Chandrasekhar, S. (1983), {\sl The Mathematical Theory of
    Black Holes} (Clarendon Press, Oxford).
\bibitem{x}McIntosh, C. B. G.  and Hickman, M. S. (1985), ``Complex Relativity and real Solutions. I. Introduction'',  {\sl Gen. Relativ. Gravit.} {\bf 17} 111-132.
\bibitem{x}Sultana, J. and Dyer, C. C.(2005),``Cosmological black holes: A black hole in the Einstein-de Sitter universe'', {\it Gen, Relatv. Gravit.} {\bf 37}, 1349-1370.
\bibitem{x}Gibbons, G. W. and Hawking, S. W. (1977), `` Cosmological event horizons, thermodynamics, and particle creation'', {\sl Phys. Rev. D} {\bf 15}, 2738-2751.
\bibitem{x}Carter, B. (1973), ``Black Hole Equilibrium States'', in {\it Black holes} edited by C Dewitt and B.C. Dewitt (New York, Gordon and Breach Science Publication).
\bibitem{x}York, J. W. (1983), ``Dynamical Origin of Black Hole radiance'', {\sl Phys. Rev. D} {\bf 28}, 2929-2945.
\bibitem{x}Choudhury, T. R. and Padmanabhan,  T. (2007) {\sl `` Concept of temperature in multi-horizon spacetimes: analysis of Schwarzschild-de Sitter metric''}, {\it Gen. Relativ Gravit.}, {\bf 39}, 1789-1811
\bibitem{x}Visser, M. (1992), ``Dirty black holes: thermodynamics and horizon structure'', {\sl Phys. Rev. D} {\bf 46}, 2445-2451.
\bibitem{x}Husain V 1996 ``Exact solutions for null collapse'', {\sl Phys. Rev.} D {\bf 53} R1759
\end{thebibliography}
\end{document}